	\documentclass[fleqn,usenatbib]{mnras}
	\PassOptionsToPackage{draft}{hyperref}

	\usepackage{newtxtext,newtxmath}

	\usepackage[T1]{fontenc}

	\DeclareRobustCommand{\VAN}[3]{#2}
	\let\VANthebibliography\thebibliography
	\def\thebibliography{\DeclareRobustCommand{\VAN}[3]{##3}\VANthebibliography}

	\usepackage{graphicx}	
	\usepackage{amsmath}	
	\usepackage{subcaption} 
	\usepackage{xspace}		

	\newcommand{\showcc}{1}  
	\newcommand{\cc}[1]{\ifnum\showcc=1 \texttt{\small \color{red}[!! #1]}\fi}    

	\newcommand{\hb}{H$\beta$\xspace}
	\newcommand{\hg}{H$\gamma$\xspace}
	\newcommand{\hei}{He\textsc{i}\xspace}  
	\newcommand{\heii}{He\textsc{ii}\xspace}
	\newcommand{\oii}{[O\textsc{ii}]\xspace}
	\newcommand{\oiii}{[O\textsc{iii}]\xspace}

	\newcommand{\neiii}{[Ne\textsc{iii}]\xspace}
	\newcommand{\ariv}{[Ar\textsc{iv}]\xspace}
	\newcommand{\fei}{[Fe\textsc{i}]\xspace}
	\newcommand{\feii}{[Fe\textsc{ii}]\xspace}
	\newcommand{\feiii}{[Fe\textsc{iii}]\xspace}
	
	\newcommand{\erg}{10$^{-16}$ erg/s/cm$^2$}

	\title[Anomalous Fe/O in XMPs]{The Mysterious Case of Iron in XMPs: Anomalous High log(Fe/O) Observed in Extremely Metal-Poor Galaxies HSCJ1631+4426 and SDSSJ0811+4730}

	\author[A. Myszka et al.]{Aaron Myszka,$^{1}$\thanks{E-mail: \href{mailto:aaronmyszka@outlook.com}{aaronmyszka@outlook.com}}
	Themiya Nanayakkara,$^{2,1}$
	Karl Glazebrook,$^{1}$
	Sarah M. Sweet,$^{3}$
	Brent Groves,$^{4,5}$ \newauthor
	Nikole M. Nielsen,$^{6,1}$
	Jarle Brinchmann,$^{7,8}$
	Yuki Isobe,$^{9,10,11}$
	Chiaki Kobayashi,$^{12}$
	Haruka Kusakabe,$^{13}$ \newauthor
	Michael V. Maseda$^{14}$
	\\$^{1}$Centre for Astrophysics and Supercomputing, Swinburne University of Technology, Hawthorn, Victoria 3122, Australia
	\\$^{2}$Sydney Institute for Astronomy, School of Physics, The University of Sydney, NSW 2006, Australia
	\\$^{3}$School of Mathematics and Physics, University of Queensland, St Lucia, Queensland 4072, Australia
	\\$^{4}$International Centre for Radio Astronomy Research, University of Western Australia, 7 Fairway, Crawley, 6009 WA, Australia
	\\$^{5}$Astronomy Australia Ltd. (Perth Office), University of Western Australia, 7 Fairway, Crawley, 6009 WA, Australia
	\\$^{6}$Homer L. Dodge Department of Physics and Astronomy, The University of Oklahoma, 440 W. Brooks St., Norman, OK 73019, USA
	\\$^{7}$Instituto de Astrofísica e Ciências do Espaço, Universidade do Porto, CAUP, Rua das Estrelas, 4150-762, Porto, Portugal
	\\$^{8}$European Southern Observatory, Karl-Schwarzschild-Straße 2, 85748 Garching, Germany
	\\$^{9}$Kavli Institute for Cosmology, University of Cambridge, Madingley Road, Cambridge, CB3 0HA, UK
	\\$^{10}$Cavendish Laboratory, University of Cambridge, 19 JJ Thomson Avenue, Cambridge, CB3 0HE, UK
	\\$^{11}$Waseda Research Institute for Science and Engineering, Faculty of Science and Engineering, Waseda University, 3-4-1, Okubo, Shinjuku, Tokyo 169-8555, Japan
	\\$^{12}$Centre for Astrophysics Research, Department of Physics, Astronomy and Mathematics, University of Hertfordshire, Hatfield AL109AB, UK
	\\$^{13}$Department of General Systems Studies, Graduate School of Arts and Sciences, The University of Tokyo, 3-8-1 Komaba, Meguro-ku, Tokyo, 153-8902, Japan
	\\$^{14}$Department of Astronomy, University of Wisconsin-Madison, Madison, WI 53706, USA
	}

	\date{Accepted XXX. Received YYY; in original form ZZZ}

	\pubyear{\the\year{}}

\begin{document}
	\label{firstpage}
	\pagerange{\pageref{firstpage}--\pageref{lastpage}}
	\maketitle



	\begin{abstract}
		We present integral field spectroscopic observations of two local extremely metal-poor galaxies (XMPs), HSCJ1631+4426 and SDSSJ0811+4730, obtained with the Keck Cosmic Web Imager (KCWI) over the wavelength range $3545-5529$ Å at a spectral sampling of 0.5 Å, capturing bright nebular emission lines essential for determining gas-phase metallicity and chemical enrichment measurements.
		Using integrated spectra, we derive oxygen abundances of $\rm12+\log(O/H)=7.079\pm0.010$ and $6.926\pm0.004$, and elevated Fe/O ratios of $\rm\log(Fe/O)=-1.57\pm0.17$ and $-1.28\pm0.07$, for HSCJ1631+4426 and SDSSJ0811+4730 respectively.
		Each galaxy is fully contained within the $\sim$8'' field of view, with 0.15'' spaxels providing spatially resolved information.
		These measurements indicate unusually efficient iron enhancement at extremely low metallicity; Fe/O ratios approach or exceed solar despite oxygen abundances of only $\sim$2\% solar, inconsistent with enrichment from core-collapse supernovae alone or delayed Type Ia supernovae given the young ages of the systems.
		Comparison with chemical evolution models suggests rare, highly energetic explosions such as bright hypernovae and/or pair-instability supernovae are likely responsible.
		Our results reinforce the growing evidence that XMPs can reflect the nucleosynthetic processes of early energetic stellar explosions, serving as local laboratories for chemical enrichment pathways prevalent in the early Universe.
		\end{abstract}





	\section{Introduction}
	\label{sec:introduction}

	The study of galaxies in the early stages of chemical enrichment reveals insight into the processes governing star formation, feedback mechanisms, and the impacts of supernovae on the surrounding gas across cosmic time.
	High $\alpha$-element abundances (e.g. O, Mg, Si, S, Ar, Ca) indicate enrichment dominated by core-collapse supernovae (CCSNe; \citealt{2013ARA&A..51..457N}), while elevated iron points to Type Ia supernovae (SNe Ia) from lower-mass binary systems \citep{2014ARA&A..52..107M}; the Fe/O ratio can be used to evaluate dominant enrichment mechanisms \citep{2011MNRAS.414.3231K,2016ApJ...826..159S,2019NatAs...3..631X}.

	Recent JWST observations have detected elevated nebular Fe/O ratios in high-redshift galaxies \citep{2024MNRAS.535..881J,2024ApJ...976..122N,2025ApJ...994...65N,2025MNRAS.540..851T,2026MNRAS.547ag123I}, in contrast to iron-deficient, $\alpha$-enhanced stellar patterns at lower redshift \citep{2019MNRAS.487.2038C,2022ApJ...925...82K,2024MNRAS.532.3102S}, highlighting the need to understand iron enrichment mechanisms across cosmic time.
	Extremely metal-poor galaxies (XMPs; $<$10\% solar metallicity) in the local universe serve as accessible analogues to these high-redshift systems \citep{2012A&A...546A.122I}, with strong nebular emission lines and low dust content enabling ideal abundance investigations \citep{Izotov2006,2010Natur.467..811C}.

	Recent studies of XMPs in the local universe have revealed unexpected chemical abundance patterns that challenge current evolution models and theory.
	The discovery and analysis of galaxy J0811+4730 by \cite{Izotov2018} as well as further works by \cite{Kojima2020,Kojima2021} and \cite{Isobe2022} reveals an anomalous large Fe/O ratio for its measure of metallicity.
	Another such case is found within galaxy J1631+4426, discovered by \cite{Kojima2020} and further explored by \cite{Kojima2021} and \cite{Isobe2022}.
	
	Exhibiting approximately 2\% solar metallicity with iron abundance levels comparable to solar, these systems show that high iron-to-oxygen ratios are not supported by any current chemical evolution models.
	Standard models involving CCSNe and SNe Ia fail to simultaneously produce the enhancement levels of both alpha-elements and iron across the short time scales experienced by these XMP galaxies \citep{Isobe2022}.
	This may indicate an influence of additional external enrichment pathways, such as feedback-driven metal redistribution or the accretion of iron-enriched circumgalactic material, though such mechanisms would require fine-tuned conditions to enhance Fe/O without correspondingly elevating the oxygen abundance, and are not supported by current observational evidence \citep{Kojima2021}.

	Both galaxies are extremely young systems, leaving insufficient time for the delayed iron enrichment from SNe Ia that would typically drive elevated Fe/O at higher metallicities.
	In young, $\alpha$-enhanced systems one would instead expect low Fe/O, yet both HSCJ1631+4426 and SDSSJ0811+4730 exhibit the opposite: Fe/O ratios approaching or exceeding solar despite oxygen abundances of only $\sim$2\% solar.
	This inversion requires a non-standard iron source capable of operating on timescales comparable to massive star evolution.

	Other potential sources of these increased iron abundances on short time scales include theoretical pair-instability supernovae (PISNe) and bright hypernovae (BrHNe).
	PISNe occur from the explosions of very massive stars ($>$140–300 M$_\odot$; \citealt{2002ApJ...567..532H}), releasing energies of $\sim 10^{53}$ erg, where high-energy photons produce electron-positron pairs, reducing the supportive radiation pressure and triggering an extreme gravitational collapse \citep{1967ApJ...148..803R,2002ApJ...567..532H}.
	BrHNe are core-collapse supernovae (CCSNe) from progenitor stars of $\sim 30-100$ M$_\odot$ with unusually high explosion energies exceeding 10$^{52}$ erg \citep{2020ApJ...892..153M}.
	Each of these events produces large amounts of $^{56}$Ni, which rapidly radioactively decays into $^{56}$Co and then into $^{56}$Fe, a stable isotope which increases the abundance of iron in these systems \citep{2008ApJ...673.1014U}.
	Furthermore, both PISNe and BrHNe events occur predominantly in low metallicities \citep{2020ApJ...892..153M} as the massive stars required can only be formed in such environments, and neither channel is captured by standard chemical evolution models.

	In this letter, we present Keck Cosmic Web Imager (KCWI) observations of galaxies HSCJ1631+4426 and SDSSJ0811+4730, two local XMPs that are among the most metal-poor galaxies known in the local universe, exhibiting extremely low O/H and anomalously high Fe/O abundances.
	Previously, analyses involving these objects have relied primarily on single-slit spectroscopy \citep{Izotov2018,Kojima2021,Isobe2022,2022MNRAS.516L..81T}, with the exception of \citet{2021PASJ...73.1631K} who presented Subaru/FOCAS IFU observations of HSCJ1631+4426 focused on its oxygen metallicity gradient.
	Here, we present KCWI IFU spectroscopy of both systems, enabling precise spatial boundary definition and higher signal-to-noise direct-method measurements of both oxygen abundance and iron enrichment.
	These observations provide crucial constraints on chemical evolution pathways in these objects and, by extension, the low-metallicity early universe counterparts, indicating a need to revise existing models of stellar feedback and metal enrichment in these primordial galaxies.
	We adopt the solar metallicity of $\rm 12+\log(O/H)=8.69$ and $\rm\log(Fe/O)$ abundance ratio of $-1.23$ \citep{2009ARA&A..47..481A,2021A&A...653A.141A}.


	\section{Observations and Data Processing}
	\label{sec:observations}

\begin{figure}
		\centering
	    \begin{subfigure}{\columnwidth}
	        \centering
	        \includegraphics[width=0.75\columnwidth]{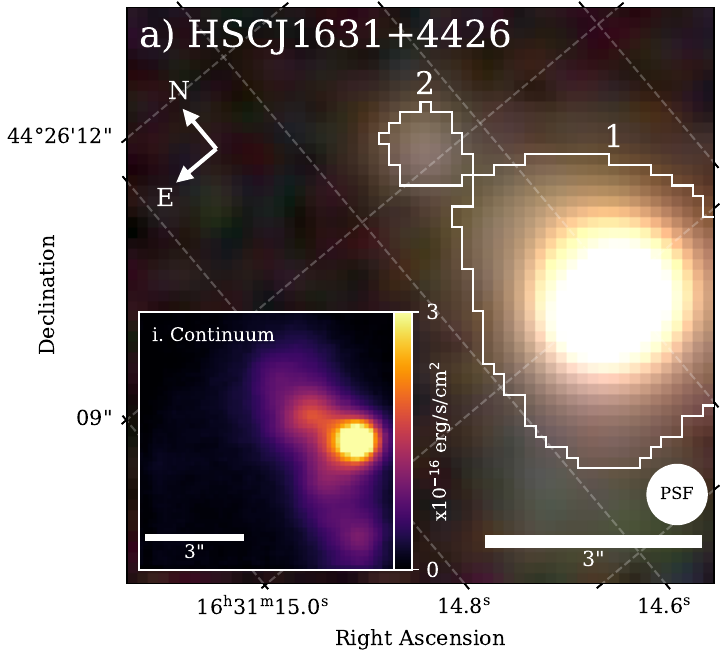}
	    	\end{subfigure}
	    \vspace{0.5em}
	    \begin{subfigure}{\columnwidth}
	        \centering
	        \includegraphics[width=0.75\columnwidth]{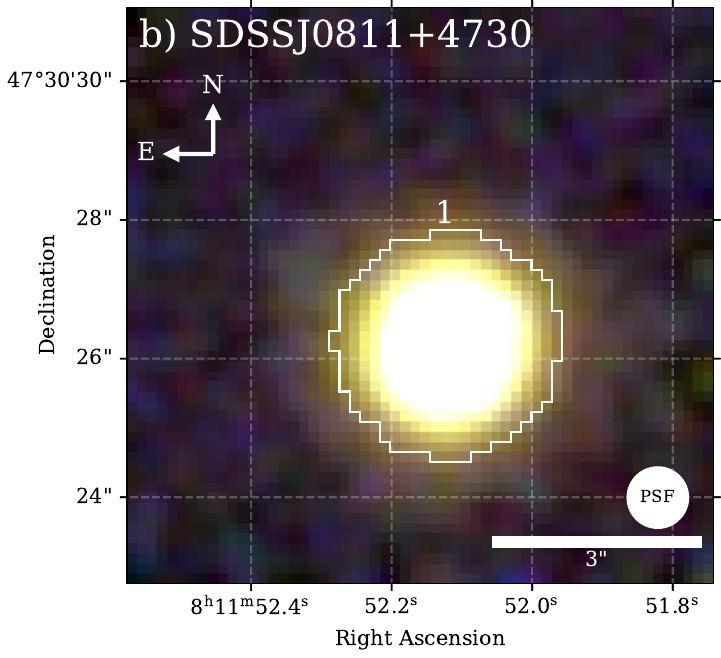}
	    	\end{subfigure}
	    \caption{
	    	Composite RGB images of (a) HSCJ1631+4426 and (b) SDSSJ0811+4730.
	    	Each image is constructed from the fitted line fluxes of \oiii $\lambda$5007 (red), \hb (green), and the combined \oii $\lambda$3726 + $\lambda$3729 doublet (blue).
	    	Galaxy boundaries identified using \texttt{astrodendro} are overlaid in white, with the individual boundaries numbered for reference.
	    	A PSF circle and 3-arcsecond scale bar are shown bottom right.
	    	An inset is included towards the lower left of panel a to display the map of the summed fitted continuum of this galaxy; such extended features are not present for SDSSJ0811+4730.
	    	}
	    \label{fig:whitelight}
		\end{figure}

	The observations of the two XMPs HSCJ1631+4426 ($z = 0.031$) and SDSSJ0811+4730 ($z = 0.044$) were obtained with the Keck Cosmic Web Imager (KCWI) on the Keck II telescope on the 25$\rm^{th}$ of March, 2022.
	All exposures were taken using the small slicer IFU covering the wavelength range $3545-5529$ Å, under seeing conditions of approximately 0.9'' for SDSSJ0811+4730 and 0.85'' for HSCJ1631+4426.
	Between sets of exposures, 90° rotations were applied to improve spatial sampling, allowing the rectangular IFU spaxels to be effectively reconstructed as square spatial elements in the final data cubes.
	A total of six 1800-second exposures were taken for SDSSJ0811+4730 and eight 1800-second exposures for HSCJ1631+4426, corresponding to total integration times of 3 hours and 4 hours respectively.
	The two galaxies in their observed field of view (FOV) are shown in Figure \ref{fig:whitelight}.

	The data were reduced and exposures combined following the methods outlined in \cite{nikki2022}. 
	We used the IDL version of the KCWI Data Reduction Pipeline (DRP; v1.2.2)\footnote{\url{https://github.com/Keck-DataReductionPipelines/KcwiDRP}} with standard settings where we skipped the sky subtraction step (\texttt{kcwi\_stage5sky}), with flux calibrations done with standard star feige56 from the KCWI DRP starlist. 
	We further reduced the data using in-house software \citep{nikki2022} to remove a residual $10\%$ illumination gradient across the FOV, producing final data cubes that are flux-calibrated, sky-subtracted using in-frame skies, and flat to within one percent. 
	We used \texttt{Montage}\footnote{\url{http://montage.ipac.caltech.edu/}} to optimally combine all exposures for a given galaxy, reproject the data onto square spaxels, and crop the data to the region over which all exposures overlap.
	The datacubes were then continuum-subtracted using the \texttt{fit\_spec} function of the \texttt{pyplatefit} package by \cite{pyplatefit}.

	Once the nebular emission was isolated, emission lines of each spectrum in each spaxel was fitted using the routine outlined within \cite{sami_zoom}, based on the \texttt{LMFIT} package by \cite{LMFIT}.
	Due to continuing difficulties with reported uncertainties within \texttt{LMFIT} (covered in \citealt{sami_zoom}), we again determined line flux uncertainties through a per-spectra RMS-based approach.
	We note that this approach does not account for Poisson noise in bright emission lines such as \hb and \oiii $\lambda$5007, and as such their uncertainties are likely underestimated. 
	However, as the faint iron lines driving our abundance measurements are dominated by continuum noise, we do not expect this to impact our derived log(Fe/O) values or Monte Carlo uncertainties.

	Prior to abundance analysis, integrated line fluxes were corrected for dust attenuation by fitting E(B-V) via least-squares minimisation over all available Balmer lines relative to theoretical Case B values \citep{1989agna.book.....O}, applying the \citet{1989ApJ...345..245C} extinction law.
	We obtain E(B-V) = $0.014 \pm 0.003$ for HSCJ1631+4426 and E(B-V) = $-0.084 \pm 0.001$ for SDSSJ0811+4730; the negative value for SDSSJ0811+4730 is clipped to zero, consistent with the negligible dust content expected at such low metallicity.
	The resulting corrections are small ($\lesssim$7\%) as expected for XMP galaxies.
	The fitted and de-reddened flux values of these emission lines are shown in Table \ref{tab:integrated_fluxes}.

	Through the initial inspection of the datacubes, it became apparent that the data for HSCJ1631+4426 included an interesting feature.
	Along with the primary body of the galaxy, there are also diffuse features extending out from the body in two directions, one towards the north and the other more south-east.
	These features are comparably quite subtle, so they can only be faintly seen in Figure \hyperref[fig:whitelight]{1a}.

	To distinguish between these features and construct integration boundaries, the \texttt{astrodendro} Python package by \cite{astrodendro} was used on the summed \hb and \oiii $\lambda$5007 flux maps. 
	Boundaries were determined by a global noise-map RMS threshold, feature contrast, and PSF size criteria.
	The resulting boundaries, including one around the brighter northern prominence of HSCJ1631+4426, are indicated in Figure \ref{fig:whitelight}.

\begin{figure*} \center 
		\includegraphics[width=0.9\textwidth]{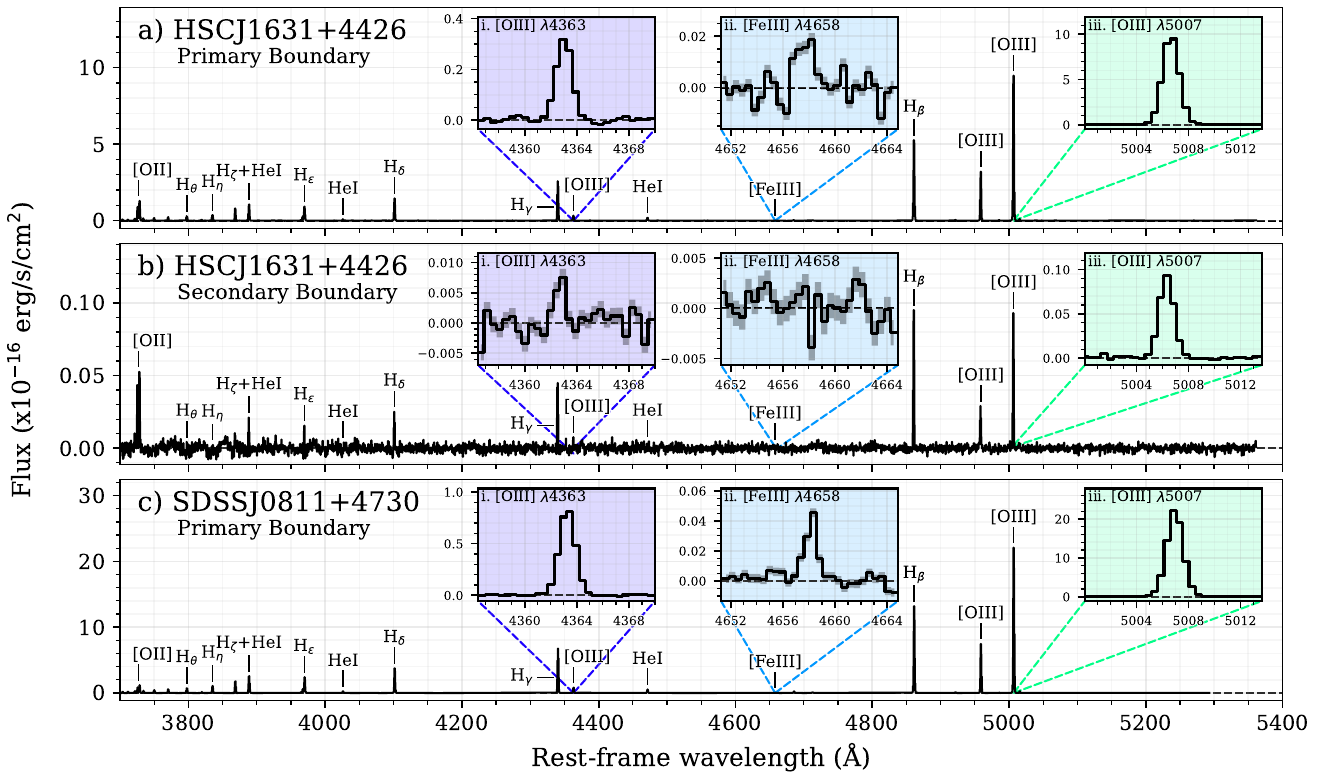}
	    \caption{
	    	The rest-frame integrated spectra of a) the primary boundary of HSCJ1631+4426, b) the secondary boundary of HSCJ1631+4426, and c) SDSSJ0811+4730.
	    	Spectra are extracted within boundaries shown in Figure \ref{fig:whitelight}, continuum-subtracted.
	    	Within each panel, insets are provided to show a more detailed view of the spectra around the positions of i. \oiii $\lambda$4363, ii. \feiii $\lambda$4658, and iii. \oiii $\lambda$5007.
	    	1$\sigma$ intervals are shown shaded in grey.
	    	} 
	    \label{fig:spec}
		\end{figure*}

	In conjunction with the spatially resolved data set, results using the fully-integrated spectra of the two galaxies were also produced.
	To integrate the three bounded sections independently, a profile-weighted combination was applied to the original datacubes using the previously fitted \oiii flux as the profile.
	These were then taken through an identical continuum fitting and subtraction process as outlined above, and then the same emission line fitting process.
	The final processed spectra can be seen in Figure \ref{fig:spec}.

\begin{table}
	\small
    \centering
    \caption{
        Fitted and de-reddened flux values for key and supplementary emission lines measured in the integrated spectra of the two XMP galaxies.
        Wavelengths are given as measured in air.
        Fluxes are shown normalised to F(\hb) = 100.
        Lines of signal-to-noise ratio less than the threshold of 3 (undetected) are given as upper limits.
        Measured line fluxes from the same spectrum (boundary) hold the same uncertainty values due to the RMS-based approach mentioned in Section \ref{sec:observations}.
        The \feii line marked with a star ($\star$) is likely blended with permitted Fe\textsc{ii} emission; non-detection of \feii $\lambda$4244 suggests the forbidden contribution is negligible.
        }
    \label{tab:integrated_fluxes}
    \setlength{\tabcolsep}{1.5pt}
    \begin{tabular}{l l r @{ $\pm$ } l r @{ $\pm$ } l r @{ $\pm$ } l}
                        &                & \multicolumn{4}{c}{\underline{HSCJ1631+4426}} 					 	& \multicolumn{2}{c}{\underline{SDSSJ0811+4730}} 	\\			
        Line 			& $\lambda$ (Å)  & \multicolumn{2}{c}{Boundary 1} 	& \multicolumn{2}{c}{Boundary 2} 	& \multicolumn{2}{c}{Boundary 1} 					\\		
        \hline
        \oii 			&      3726.032  &  15.35 & 0.11 					& 51.06 & 2.58 						&   5.91 & 0.04           							\\
        \oii   			&      3728.815  &  22.42 & 0.11 					& 68.86 & 2.58						&   7.90 & 0.04           							\\
        \hei   			& 	   3819.6074 &   0.88 & 0.11 					& \multicolumn{2}{c}{< 7.74}		&   0.96 & 0.04           							\\
        \neiii          & 	   3868.760  &  14.00 & 0.11 					& 11.55 & 2.58						&  12.87 & 0.04          							\\
       	\hei 			& 	   3888.648  &  13.88 & 0.11 					& 18.44 & 2.58 						&  14.53 & 0.04          							\\
        \hei 			& 	   3964.7291 &   0.77 & 0.11 					& \multicolumn{2}{c}{< 7.74}		&   0.80 & 0.04           							\\
        \neiii 			& 	   3967.470  &   5.09 & 0.11 					& \multicolumn{2}{c}{< 7.74}		&   4.55 & 0.04           							\\
        H$\epsilon$     & 	   3970.079  &  16.19 & 0.11 					& 14.28 & 2.58 						&  17.04 & 0.04          							\\
        \hei 			& 	   4026.190  &   1.65 & 0.11 					& \multicolumn{2}{c}{< 7.74}		&   1.88 & 0.04           							\\
        H$\delta$       & 	   4101.742  &  26.00 & 0.11 					& 24.98 & 2.58 						&  27.29 & 0.04          							\\
        \hei 			& 	   4143.761  & \multicolumn{2}{c}{< 0.33} 		& \multicolumn{2}{c}{< 7.74}		&   0.31 & 0.04           							\\
        \hg             & 	   4340.471  &  46.83 & 0.11 					& 50.47 & 2.58 						&  49.65 & 0.04          							\\
        \feii$^\star$ 		& 	   4358.162  & \multicolumn{2}{c}{< 0.33} 		& \multicolumn{2}{c}{< 7.74}		&   0.15 & 0.04										\\
        \oiii  			&      4363.210  &   5.63 & 0.11 					& \multicolumn{2}{c}{< 7.74}		&   6.10 & 0.04           							\\
        \hei   			& 	   4387.929  &   0.37 & 0.11 					& \multicolumn{2}{c}{< 7.74} 		&   0.47 & 0.04           							\\
        \hei   			& 	   4471.479  &   3.45 & 0.11 					& \multicolumn{2}{c}{< 7.74} 		&   3.96 & 0.04           							\\
        \feiii 			&      4658.050  &   0.34 & 0.11 					& \multicolumn{2}{c}{< 7.74}		&   0.27 & 0.04           							\\
        \heii  			& 	   4685.710  &   0.72 & 0.11 					& \multicolumn{2}{c}{< 7.74}		&   1.83 & 0.04           							\\
        \ariv  			& 	   4711.260  &   0.35 & 0.11 					& \multicolumn{2}{c}{< 7.74}		&   0.38 & 0.04           							\\
        \hei  			& 	   4713.1457 &   0.46 & 0.11 					& \multicolumn{2}{c}{< 7.74}		&   0.64 & 0.04           							\\
        \ariv  			& 	   4740.120  & \multicolumn{2}{c}{< 0.33} 		& \multicolumn{2}{c}{< 7.74}		&   0.31 & 0.04           							\\
        \hb         	&      4861.333  & 100.00 & 0.11 					& 100.00 & 2.58 					& 100.00 & 0.04         							\\
        \hei   			& 	   4921.9313 &   0.87 & 0.11 					& \multicolumn{2}{c}{< 7.74}		&   0.91 & 0.04           							\\
        \oiii  			&      4958.911  &  56.96 & 0.11 					& 27.76  & 2.58						&  54.21 & 0.04          							\\
        \fei   			& 	   4985.9838 &   0.66 & 0.11 					& \multicolumn{2}{c}{< 7.74}		&   0.31 & 0.04           							\\
        \oiii  			&      5006.843  & 171.98 & 0.11 					& 87.27  & 2.58						& 162.23 & 0.04         							\\
        \hei   			& 	   5015.6783 &   1.76 & 0.11 					& \multicolumn{2}{c}{< 7.74}		&   1.81 & 0.04           							\\
        \hline
        \multicolumn{2}{l}{F(\hb), \erg} & 9.941  & 0.010                   & 0.285  & 0.007                    & 22.849 & 0.010                                    \\			
        \end{tabular}
    \end{table}


	\section{Results}
	\label{sec:results}

	We measure elemental abundances through the use of the \texttt{PyNeb} Python package by \citet[v1.1.22]{pyneb}, adopting a two-zone temperature scheme with separate electron temperatures for the O$^+$ and O$^{2+}$ ionisation zones.
	We first evaluate density and \oiii electron temperature iteratively through the use of the \texttt{getTemDen} procedure, using \oii $\rm F(\lambda3726)/F(\lambda3729)$ as the density line ratio and \oiii $\rm F(\lambda4959 + \lambda5007)/F(\lambda4363)$ as the temperature ratio.
	The initial temperature is set to $10^4$ K, and the initial density is set to 100 cm$^{-3}$; these quickly converge to the reported values within a small number of iterations ($<5$).
	The \oii temperature is then evaluated using the empirical $\rm T_e \left( \oii \right) = 0.7 \; T_e \left( \oiii \right) + 3000 \; K$, as verified by \cite{1986MNRAS.223..811C} and \cite{1992AJ....103.1330G}.

	The ionic abundances of O$^+$ and O$^{2+}$ are then calculated using the \texttt{getIonAbundance} procedure within \texttt{PyNeb}.
	The O$^+$ abundance was determined using the combined \oii $\lambda\lambda$3726,9 flux relative to the H$\beta$ flux, while the O$^{2+}$ abundance used the relative \oiii $\lambda$4959 + $\lambda$5007 flux.
	These abundance values are then combined under the assumption $\rm O/H = O^+/H^+ + O^{2+}/H^+$ to calculate the final oxygen abundance.

	Measurements within the second boundary of HSCJ1631+4426 could not be reliably determined because the required emission lines were unable to be detected in its integrated spectrum.
	Therefore, this region is excluded from the study. 
	Results from the primary boundary of this galaxy are hereafter referred to simply as HSCJ1631+4426.

	We determine the metallicity within HSCJ1631+4426 to be $\rm12+\log(O/H)=7.079\pm0.010$ and that of SDSSJ0811+4730 to be $\rm12+\log(O/H)=6.926\pm0.004$, broadly consistent with, though not within $1\sigma$ of, previous literature values of $6.90\pm0.03$ \citep{Kojima2020} and $6.98\pm0.02$ \citep{Izotov2018} respectively.

	The ionic abundance of Fe$^{2+}$ is similarly determined using the \texttt{getIonAbundance} procedure, using the measured \feiii $\lambda$4658 flux relative to H$\beta$.
	A value for the total iron abundance is then obtained through the \texttt{getElemAbundance} step, taking the O$^+$ and O$^{2+}$ abundances in conjunction with the low-metallicity ionisation correction factor (ICF) of \cite{Izotov2006}.

	Through this method, we obtain a value of $\rm \log(Fe/O)=-1.57\pm0.17$ for HSCJ1631+4426 and a value of $\rm \log(Fe/O)=-1.28\pm0.07$ for SDSSJ0811+4730, consistent with but significantly more precise than previous measurements of $-1.25^{+0.17}_{-(0.31+0.22)}$ and $-1.06^{+0.09}_{-(0.09+0.22)}$ \citep{Kojima2021}, reflecting our more secure \feiii $\lambda$4658 detections.

	The uncertainties on the metallicity and iron enrichment values are determined by performing 1000 Monte-Carlo iterations, altering line flux measurements by sampling a normal distribution of standard deviation equal to the respective line flux uncertainties each time.
	We also determine elemental abundances adopting a second ionisation correction model, that of \cite{2005ApJ...626..900R}.
	This model applies a different ionisation correction scheme, and has been shown to produce values approximately 0.2 dex lower \citep{Kojima2021,Isobe2022}.
	Through this process we find iron enrichments of $-1.80\pm0.16$ for HSCJ1631+4426 and $-1.55\pm0.08$ for SDSSJ0811+4730, corresponding to decreases in $\rm \log(Fe/O)$ of 0.23 and 0.27, respectively, in agreement with previous literature.
	These differences are treated as additional systematic uncertainties and discussed further in Section \ref{sec:discussion}.

	We also test the sensitivity of the derived $\rm\log(Fe/O)$ to the choice of \feiii atomic data by repeating the abundance calculation across all six datasets available in \texttt{PyNeb}\footnote{The six \feiii atomic datasets available in \texttt{PyNeb} v1.1.22 are from \citet{2014ApJ...785...99B} (144 levels, default), \citet{2014ApJ...785...99B} (34 levels), \citet{2009ADNDT..95..184D}, \citet{1996A&AS..119..509N}, \citet{1996A&AS..116..573Q}, and \citet{1996A&AS..116..573Q} combined with \citet{2000A&A...361..977J}, all revised by \citet{2023Atoms..11...63M}.}, finding that the derived values range from $-1.62$ to $-1.52$ (0.10 dex) for HSCJ1631+4426 and $-1.35$ to $-1.22$ (0.13 dex) for SDSSJ0811+4730, with the adopted default dataset sitting near the centre of each distribution.

	Integrated $\rm 12 + \log(O/H)$ metallicity and $\rm \log(Fe/O)$ iron enrichment values are shown, together with the derived intermediate quantities and additional sources of uncertainty, in Table \ref{tab:int_properties}.

	\begin{table}
		\small
	    \centering
	    \caption{
	    	Integrated properties of the primary boundary of HSCJ1631+4426 and SDSSJ0811+4730.
	        The properties of the secondary boundary of HSCJ1631+4426 are not given as none were measurable.
	        Reported uncertainties on $\rm log(Fe/O)$ reflect Monte Carlo propagation of line flux uncertainties only.
			$\Delta\rm log(Fe/O)_\mathrm{ICF}$ and $\Delta\rm log(Fe/O)_\mathrm{atom}$ give the additional systematic offsets from the choice of ionisation correction factor and \feiii atomic data respectively, as discussed in Section \ref{sec:discussion}.
	        }
	    	\begin{tabular}{l r@{$\,\pm\,$}l r@{$\,\pm\,$}l}
			    Property                                    & \multicolumn{2}{c}{HSCJ1631+4426}     & \multicolumn{2}{c}{SDSSJ0811+4730}    \\
			    \hline \\ [-10pt]
			    EW$_0$(H$\beta$)                            & 121.9     & 0.6       & 271.3     & 1.3       \\ [5pt]
			    T$\rm_e$(\oiii), K                          & 19729     & 221       & 21522     & 107       \\
			    T$\rm_e$(\oii), K                           & 16810     & 155       & 18066     & 75        \\
			    n$\rm_e$, cm$^{-3}$                         & 17.63     & 9.08      & 129.64    & 13.00     \\ [5pt]
			    O$^+$/H$^+$ $(\times10^{-6})$               & 2.19      & 0.06      & 0.674     & 0.008     \\
			    O$^{2+}$/H$^+$ $(\times10^{-6})$            & 9.80      & 0.23      & 7.75      & 0.08      \\
			    O/H $(\times10^{-5})$                       & 1.20      & 0.03      & 0.843     & 0.008     \\ [5pt]
			    12 + log(O/H)                               & 7.079     & 0.010     & 6.926     & 0.004     \\ [5pt]
			    Fe$^{2+}$/H$^+$ $(\times10^{-8})$           & 4.29      & 1.36      & 2.56      & 0.42      \\
			    Fe/H $(\times10^{-7})$                      & 3.19      & 1.01      & 4.40      & 0.73      \\ [5pt]
			    $\rm log(Fe/O)$                             & $-1.57$   & 0.17      & $-1.28$   & 0.07      \\ [5pt]
	    		$\Delta\rm log(Fe/O)_\mathrm{ICF}$      & \multicolumn{2}{c}{\hspace{0.76cm}$-0.23\phantom{\pm0.00}$}  & \multicolumn{2}{c}{\hspace{0.8cm}$-0.27\phantom{\pm0.00}$}  \\
	        	$\Delta\rm log(Fe/O)_\mathrm{atom}$     & \multicolumn{2}{c}{\hspace{0.76cm}$\pm0.10\phantom{-0.00}$}  & \multicolumn{2}{c}{\hspace{0.8cm}$\pm0.13\phantom{-0.00}$}  \\
			    \end{tabular}
	    \label{tab:int_properties}
	    \end{table}

	If we neglect to separate the extended features from the primary structure of HSCJ1631+4426, combining the light from across the whole field of view, we approach near identical metallicity and iron enrichment values of $\rm 12+\log(O/H)=7.08 \pm 0.02$ and $\rm \log(Fe/O)=-1.57 \pm 0.19$.
	This indicates that any differences to the line ratios and abundances within the extended features of this galaxy are completely dominated by the bright emissions contained within its primary boundary and are therefore unmeasurable in the data.

	We measure the equivalent width of H$\beta$, EW$_0$(\hb), from the continuum fit in spectral windows either side of the \hb line (4825--4845 Å and 4880--4900 Å), obtaining $121.9 \pm 0.6$ Å and $271.3 \pm 1.3$ Å for HSCJ1631+4426 and SDSSJ0811+4730 respectively. 

 
	\section{Discussion}
	\label{sec:discussion}

	The derived oxygen abundances confirm that both HSCJ1631+4426 and SDSSJ0811+4730 are among the most chemically pristine star-forming galaxies in the known local Universe; both fall significantly below 10\% Z$_\odot$ to be classed as XMPs.
	Coupled with the elevated Fe/O ratios measured in both systems, this strongly suggests a non-standard enrichment pathway not captured by conventional chemical evolution models.
	
	The contributions of Type Ia supernovae (SNe Ia) drive an increase in $\rm \log(Fe/O)$ at metallicities of $\rm 12+\log(O/H) > 8$ \citep{2006MNRAS.367.1181B}, with the accumulation of iron below this threshold owing primarily to core-collapse supernovae (CCSNe, \citealt{Kojima2021}).
	When compared to Galactic stellar abundances, the two galaxies exhibit a clear divergence from the Milky Way Fe/O evolutionary track, being well below $\rm 12+\log(O/H)=8$ with very high $\rm \log(Fe/O)$.

	\begin{figure} \center 
		\includegraphics[width=\columnwidth]{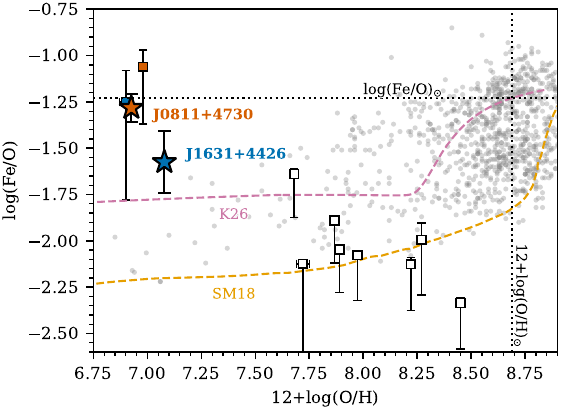}
	    \caption{
			A comparison of Fe/O and $\rm 12+\log(O/H)$ data, adapted from \protect\citet[K21]{Kojima2021}, with our results for the primary boundary of HSCJ1631+4426 and SDSSJ0811+4730 shown as coloured stars.
			Error bars on the star symbols indicate Monte Carlo uncertainties; systematic uncertainties from the ICF and \feiii atomic data choices are discussed in Section \ref{sec:results}.
			K21 sample galaxies are shown as square points, with HSCJ1631+4426 and SDSSJ0811+4730 in their respective colours; stellar data from \protect\cite{2014A&A...562A..71B}, \protect\cite{2004A&A...416.1117C}, \protect\cite{2003A&A...404..187G}, \protect\cite{1993A&A...275..101E}, and \protect\cite{2003MNRAS.340..304R} are shown as circles.
			Dashed lines show the Milky Way chemical evolution models of \protect\cite{2018ApJ...852..101S} ($9 - 100$ M$_\odot$ gas enrichment) and an updated \protect\cite{2020ApJ...900..179K} model (K26) with a flattened Population III IMF ($10 - 280$ M$_\odot$, $x=0$ vs.\ standard Salpeter $x=1.35$) enabling PISNe.
			Uncertainties in $\rm 12+\log(O/H)$ are too small to be seen under point markers; see Table \ref{tab:int_properties}.
	    	}
	    \label{fig:logFeO_Z}
		\end{figure}

	Figure \ref{fig:logFeO_Z} shows the positions of the two XMP galaxies in metallicity-$\rm \log(Fe/O)$ space, highlighting their anomalously high iron enrichment.

	\cite{Kojima2021} investigate whether the elevated $\rm \log(Fe/O)$ ratios could be artificially inflated by hard extreme-ultraviolet radiation, shock heating, or contamination from C\textsc{iv} $\lambda$4659, and rule out each of these effects as a dominant contributor.
	The authors also find that the preferential depletion of iron into dust relative to oxygen dominates as metallicity increases, producing the negative correlation between $\rm \log(Fe/O)$ and $\rm 12+\log(O/H)$ visible in the K21 sample in Figure \ref{fig:logFeO_Z} (square points).
	However, such dust formation is negligible at the extremely low metallicities of HSCJ1631+4426 and SDSSJ0811+4730; this effect cannot explain their elevated iron enhancements.
	
	\cite{Isobe2022} additionally exclude the contributions of SNe Ia to the iron enrichment of SDSSJ0811+4730, due to its approximate age of only $\sim$10 Myr, as well as indicating HSCJ1631+4426 cannot be consistent with the MW model to 1$\sigma$. 
	Their measurements of the N/O ratio for each galaxy also support their conclusions.
	Based on these constraints, the authors identify enrichment from BrHNe and/or PISNe as the most plausible mechanisms for producing the elevated iron enrichment.
	We note that the previous \feiii $\lambda$4658 detection in HSCJ1631+4426 was marginal at $2.4\sigma$ \citep{Kojima2021}; our KCWI data improves this to $3.1\sigma$ and $6.8\sigma$ for HSCJ1631+4426 and SDSSJ0811+4730 respectively, lending greater confidence to the derived Fe/O values.

	\begin{figure} \center 
		\includegraphics[width=\columnwidth]{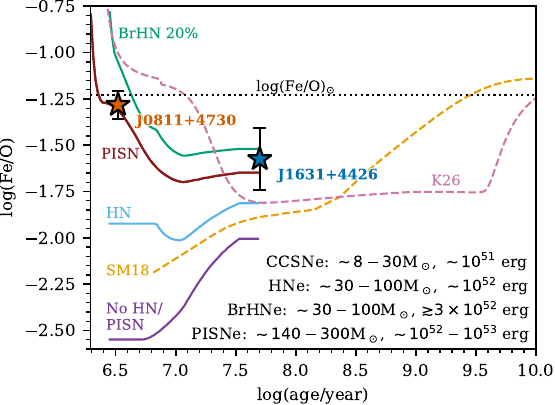}
	    \caption{
			Fe/O-age evolutionary tracks, adapted from \protect\cite{Isobe2022}, with our results for HSCJ1631+4426 and SDSSJ0811+4730 overlaid.
			Solid lines indicate theoretical chemical evolution model pathways \protect\citep{Isobe2022} for a 20\% proportion of bright hypernovae (BrHN 20\%), pair-instability supernovae (PISN), entirely hypernovae (HN), or no hypernovae/PISNe (No HN/PISN).
			Dashed lines show the Milky Way Fe/O models of \protect\cite{2018ApJ...852..101S} ($9-100$ M$_\odot$ gas enrichment) and the updated \protect\cite{2020ApJ...900..179K} model (K26), which flattens the Population III IMF ($10-280$ M$_\odot$, $x=0$ vs.\ Salpeter $x=1.35$) to enable PISNe.
			Error bars indicate Monte Carlo uncertainties; ICF and \feiii atomic data systematics are discussed in Section \ref{sec:results}.
			The solar log(Fe/O) is shown as the horizontal dotted line.
			Progenitor mass and explosion-energy ranges from \protect\citet{2002ApJ...567..532H} and \protect\cite{Isobe2022}; HNe and BrHNe share similar progenitor mass ranges but differ in explosion energy, with BrHNe representing the higher-energy subset.
	    	}
	    \label{fig:FeO_evo}
		\end{figure}

	Figure \ref{fig:FeO_evo} displays several Fe/O-age evolutionary tracks with the positions of the two XMP galaxies overlaid.
	Here, we combine the iron enrichment values determined through this work with the 50 Myr age of HSCJ1631+4426 reported by \cite{Kojima2020} and the age of SDSSJ0811+4730 by \cite{Izotov2018} of 3.3 Myr.
	This comparison indicates that for such elevated iron enrichment to occur, current models predict that contributions from the highly energetic bright hypernovae (BrHNe) and/or pair-instability supernovae (PISNe) are required at the young ages of these systems.
	Independent EW$_0$(H$\beta$) age estimates from \texttt{pyStarburst99} \citep{pyStarburst99} yield $\sim$2.8 Myr for SDSSJ0811+4730 and $\sim$4.1 Myr for HSCJ1631+4426, significantly younger than the $\sim$50 Myr of \cite{Kojima2020} derived via SED fitting, likely reflecting the greater sensitivity of the EW method to the current starburst rather than the integrated stellar population.
	Regardless of which estimate is adopted, both ages are well below the $\sim$1 Gyr delay time for significant SNe Ia iron enrichment \citep{2014ARA&A..52..107M}, reinforcing that the observed Fe/O excess cannot be attributed to delayed iron production.

	We note that both galaxies have stellar masses of $\log(M_\star/M_\odot) \lesssim 6.3$ \citep{Izotov2018,Kojima2020}, raising the question of whether the high-mass stellar tail of the IMF is sufficiently populated to produce PISNe or BrHNe in a statistical sense.
	We estimate that for a \cite{2003PASP..115..763C} IMF with an upper mass cut-off at $300\,\rm M_\odot$, stars above $140\,\rm M_\odot$ constitute $\sim$0.01\% of all stars, such that a $10^6\,\rm M_\odot$ burst is expected to produce $\sim$170 such progenitors over its lifetime, sufficient to drive significant iron enrichment through multiple events rather than a single statistical fluctuation.

	While the results favour enrichment pathways involving PISNe and/or BrHNe, we emphasise that iron abundance measurements in XMPs carry substantial systematic uncertainty that the literature has yet to fully resolve.
	The dominant sources are the choice of ionisation correction factor, which alone introduces a systematic uncertainty of $\sim0.2-0.3$ dex seen in this study and the previous by \cite{Isobe2022}, the choice of \feiii atomic data, which contributes a further $\sim0.10-0.13$ dex as quantified in Section \ref{sec:results}, and the reliance on a single \feiii emission line.
	We also note that the \cite{Izotov2006} ICF was calibrated under standard star-forming conditions, and harder ionising radiation in these systems could in principle affect the iron ICF more than other abundance ratios; however, comparing the ionisation parameter proxy \oiii/\oii against $\rm\log(Fe/O)$ across our sample and the broader XMP sample of \cite{Kojima2021} reveals no significant correlation (Spearman $\rho=0.17$, $p=0.61$), suggesting this is not the dominant driver.
	Despite these uncertainties, the persistence of elevated $\rm \log(Fe/O)$ across multiple independent studies, ionisation models, and observational datasets demonstrates that the conclusion of anomalously efficient iron production at extremely low metallicity is robust.
	Firmly establishing the enrichment mechanism will ultimately require both improved atomic data for Fe$^{2+}$ and observations of additional abundance diagnostics.
	
	In a sample of local XMPs, \cite{2024ApJ...962...50W} demonstrates that Ar/O and S/O ratios fall well below the predictions of PISN chemical evolution models, favouring enrichment by CCSNe/HNe with mixing and fallback over PISNe, despite those same galaxies exhibiting elevated Fe/O.
	Future red-optical observations extending to the [Ar\textsc{iii}] $\lambda$7136 and [S\textsc{ii}]/[S\textsc{iii}] lines in HSCJ1631+4426 and SDSSJ0811+4730 would therefore be valuable for determining whether PISNe are truly required to explain the observed iron enrichment, or whether BrHNe alone can suffice.



	\section{Summary and Conclusions}
	\label{sec:conclusion}

	We present spatially resolved KCWI spectroscopy of the extremely metal-poor galaxies HSCJ1631+4426 and SDSSJ0811+4730 ($Z < 2.5\%$ Z$_\odot$), which exhibit unusually high iron enrichment relative to oxygen despite their chemically primitive nature.
	Integrating the KCWI spectra within well-defined spatial boundaries, we measure over 40 nebular emission lines to derive electron temperatures, densities, and gas-phase abundances, obtaining $\rm12+\log(O/H)=7.079\pm0.010$ and $\rm \log(Fe/O)=-1.57\pm0.17$ for HSCJ1631+4426, and $\rm12+\log(O/H)=6.926\pm0.004$ and $\rm \log(Fe/O)=-1.28\pm0.07$ for SDSSJ0811+4730.

	Despite their extremely low oxygen abundances, both galaxies display Fe/O ratios comparable to Milky Way stars at much higher metallicities, inconsistent with standard CCSNe yields or delayed SNe Ia enrichment given their young ages; comparison with chemical evolution models instead requires contributions from rare, highly energetic explosions such as PISNe and/or BrHNe.

	While iron abundances are sensitive to the choice of ionisation correction and \feiii atomic data, the persistence of elevated Fe/O across multiple studies and datasets confirms the conclusion of unusually efficient iron production at extremely low metallicity is robust; the derived abundances are dominated by the brightest emission, with no measurable contribution from lower-surface-brightness extended features.

	These results strengthen the growing evidence that early chemical enrichment in extremely metal-poor galaxies may be driven by rare, highly energetic stellar explosions, with HSCJ1631+4426 and SDSSJ0811+4730 providing valuable local environments for constraining nucleosynthetic pathways that likely operated in the early Universe. 
	The spatially resolved KCWI data further offers the opportunity for future full spectral fitting to characterise stellar populations, kinematics, and ISM conditions across these systems.



	\section*{Acknowledgements}

	AM is grateful for the support through an Australian Government Research Training Program Scholarship (RTPS).
	This research was also supported by the Australian Research Council Centre of Excellence for All Sky Astrophysics in 3 Dimensions (ASTRO 3D), through project number CE170100013.
	TN and KG acknowledge support from Australian Research Council Laureate Fellowship FL180100060.
	TN acknowledges support from Australian Research Council Discovery Project DP230103161.
	SMS acknowledges funding from the Australian Research Council (DE220100003).
	JB acknowledges support by FCT through national funds (UID/FIS/04434/2013) and by FEDER through COMPETE2020 (POCI-01-0145-FEDER-007672).
	YI is supported by JSPS KAKENHI Grant No. 24KJ0202.
	MVM is supported by the National Science Foundation via AAG grant 2205519.
	We thank the staff located at the W. M. Keck Observatory for their assistance during the observations.
	The observations reported here were obtained under Keck program 2022A\_W226.

	Some of the data presented herein were obtained at Keck Observatory, which is a private 501(c)3 non-profit organisation operated as a scientific partnership among the California Institute of Technology, the University of California, and the National Aeronautics and Space Administration. The Observatory was made possible by the generous financial support of the W. M. Keck Foundation. 
	The authors wish to recognise and acknowledge the very significant cultural role and reverence that the summit of Maunakea has always had within the Native Hawaiian community. We are most fortunate to have the opportunity to conduct observations from this mountain. 

	\textit{Facility:} Keck:II (KCWI)



	\section*{Data Availability}

	The data underlying this article will be shared on reasonable request to the corresponding author.



	\bibliographystyle{mnras}
	\bibliography{bibliography} 






	% ...


	\bsp	
	\label{lastpage}
	\end{document}